\documentclass[useAMS,usenatbib]{mn2e}

\usepackage{natbib}
\usepackage{graphicx}
\newcommand{\mnras}{MNRAS}
\newcommand{\apj}{ApJ}
\newcommand{\apjl}{ApJL}
\newcommand{\aap}{A\&A}

\newcommand{\eqb}{\begin{equation}}
\newcommand{\eqe}{\end{equation}}

\begin{document}

\title{Highly magnetized region in pulsar wind nebulae and origin of the Crab gamma-ray flares}
\author[Y.E.Lyubarsky]{Y.E.Lyubarsky\\
Physics Department, Ben-Gurion University, P.O.B. 653, Beer-Sheva
84105, Israel; e-mail: lyub@bgu.ac.il}
\date{Received/Accepted}
\maketitle
\begin{abstract}
The recently discovered gamma-ray flares  from the Crab nebula are generally attributed to the magnetic energy release in a highly magnetized region within the nebula. I argue that such a region naturally arises in the polar region of the inner nebula.
In pulsar winds, efficient dissipation of the Poynting flux into the plasma energy occur only in the equatorial belt where the energy is predominantly transferred by alternating fields.
At high latitudes, the pulsar wind remains highly magnetized therefore the termination shock in the polar region is weak and the postshock flow remains relativistic. I study the structure of this flow and show that the flow at first expands and decelerates and then it converges and accelerates. In the converging part of the flow, the kink instability triggers the magnetic dissipation. The energy release zone occurs at the base of the observed jet. A specific turbulence of relativistically shrinking magnetic loops efficiently accelerates particles so that the synchrotron emission in the hundreds MeV band, both persistent and flaring, comes from this site.
\end{abstract}

\section{Introduction}

The unexpected discovery of strong, short gamma-ray flares from the Crab nebula \citep{Tavani11,Abdo11} called attention to the processes that occur in pulsar wind nebulae once again. During these events, the flux in a few hundred MeV band grows significantly and varies at a time-scale of a half of a day. The observed properties of the flares
place severe limits on possible models. The short time-scale implies a very compact emission region; the high energy of emitted photons is an evidence for an extremely efficient acceleration process. The underlying physical mechanism  is generally attributed \citep{Bednarek_Idec11,Uzdensky11,Cerutti12,Clausen_Lyutikov12,Sturrock_Aschwanden12} to a rapid magnetic energy release via, e.g., reconnection, which assumes a magnetically dominated region within the nebula. 

Even though the pulsar winds are highly magnetized, it is widely believed that at the termination shock, the magnetization is already extremely small. This conclusion is based on the results of spherically or axisymmetric models \citep{Rees_Gunn94,Kennel_Coroniti94,Komissarov_Lyubarsky04,DelZanna04}, which demonstrate that the magnetic hoop stress could easily distort the nebula beyond the observational limits. \citet{Begelman98} argued that beyond the shock, the magnetic field is kink unstable and would not necessarily pinch the flow as much as would otherwise be supposed; this conjecture is supported by numerical simulations \citep{Mizuno11}. In this case the wind magnetization might not be so unreasonably small as spherically and axi- symmetric models suggest.

According to the pulsar wind theory, the Poynting flux could be converted to the plasma energy only via dissipation of variable fields (see, e.g., reviews by \citet{Arons07} and \citet{Kirk_Lyubarsky_Petri09}). In the equatorial belt, the pulsar magnetic field changes polarity every half of period so that a striped wind is formed \citep{Michel71,Coroniti90}. The alternating field decays either already in the wind or at the termination shock \citep{Lyubarsky_Kirk01,Kirk_Olaf03,Lyubarsky03,Petri_Lyubarsky07,zenitani_hoshino07,Sironi_Spitkovsky11} therefore in the equatorial belt, a weakly magnetized plasma is injected into the nebula; it is this plasma that forms a bright X-ray torus. At high latitudes, the magnetic field does not change sign. The obliquely rotating magnetosphere excites fast magnetosonic waves in this part of the wind \citep{Bogovalov01}, which decay via non-linear steepening and formation of multiple shocks \citep{Lyubarsky03fms}, but the fraction of the energy transferred by the waves is not large therefore even after the waves decay, the flow remains highly magnetized. In this paper, I study the fate of the highly magnetized plasma injected into the nebula at high latitudes.

The paper is organized as follows. In the next section, I analyze the lateral distribution of the pulsar wind parameters. In sect. 3, the equations governing the high latitude flow in the nebula are derived.  In sect. 4, the structure of the flow is found. In sect. 5, I shortly discuss stability of the flow and mechanisms of the magnetic energy release and particle acceleration. Conclusions are presented in sect. 6.

\section{Preliminary considerations}
The pulsar wind is highly anisotropic; one can conveniently adopt the lateral distribution of the Poynting flux in the monopole wind \citep{Michel73,Bogovalov99}
 \eqb
F=F_0\sin^2\theta,
 \label{energy_flux}\eqe
where $F$ is the energy injected per unit time and unit solid angle, $\theta$ the polar angle. Most of the energy is transferred in the equatorial belt where the flow has a structure of the striped wind. If the pulsar inclination angle is $\alpha$, the  striped wind zone is formed at the polar angles exceeding
 \eqb
\theta_0=\frac{\pi}2-\alpha.
 \eqe
Since the alternating field decays, a weakly magnetized plasma is injected into the nebula in the equatorial belt.

At high latitudes,  $\theta<\theta_0$, the magnetic field does not change polarity; the obliquely rotating magnetosphere just excites fast magnetosonic waves  propagating outwards \citep{Bogovalov01}. These waves eventually decay via non-linear steepening and formation of multiple shocks \citep{Lyubarsky03fms} therefore within the nebula, only the mean field survives, the energy of variable fields being converted into the plasma energy. One can calculate the final (after the waves decay) wind magnetization assuming that the wave has a sine-like form such that the magnetic structure is locally presented as
 \eqb
B=B_0\left[1+\xi\sin(\Omega r)\right],
 \eqe
where $\Omega$ is the pulsar angular velocity, $\xi<1$ the relative amplitude of the wave, $B_0$ the slowly varying ($\propto 1/r$) mean field. Throughout the paper, the speed of light is taken to be unity. In the relativistic wind, the total Poynting flux is
\eqb
F=\frac{\langle B^2\rangle}{4\pi}=\frac{B^2_0}{4\pi}\left(1+\frac{\xi^2}2\right).
\eqe
After the waves decay, the Poynting flux decreases to
 \eqb
\widetilde{F}=\frac{\langle B\rangle^2}{4\pi}=\frac{B^2_0}{4\pi},
 \eqe
the residual being converted into the plasma energy. The plasma magnetization, defined as the ratio of the Poynting to the plasma energy fluxes, after the wave decay is found as (one can safely neglect the initial plasma energy)
 \eqb
\sigma=\frac{\widetilde{F}}{F-\widetilde{F}}=\frac 2{\xi^2}.
 \eqe
One sees that after the waves decay, the wind magnetization at high latitudes, $\theta<\theta_0$, remains large, $\sigma>2$.

The lateral dependance of the wind magnetization is very important because it leads to the observed disc-jet dichotomy in PWNe \citep{Lyubarsky02}: the disc is formed by the weakly magnetized plasma from the striped part of the wind whereas the high latitude flow is collimated by the magnetic hoop stress into a jet-like feature. In simulations of the PWN structure \citep{Komissarov_Lyubarsky04,DelZanna04,DelZanna06}, the magnetization was chosen to be zero at the equator and growing with latitude, as it should be, but even at high latitudes, $\sigma$ was taken to be well below unity. The reason was that high and even moderate $\sigma$ flows are subject to numerical instabilities and moreover in axisymmetric simulations, one has to reduce magnetization in order to suppress inappropriately strong elongation of the nebula. The study of the PWN structure with realistic magnetization requires three-dimensional simulations so that the kink instability \citep{Begelman98,Mizuno11} could be taken into account. Recently \citet{Komissarov12} also pointed out that the residual pulsar wind magnetization is larger than the available simulations assume, especially at high latitudes, so that magnetic dissipation via three-dimensional effects is crucially important.

As a preliminary step, 
I consider a highly magnetized conical flow injected into a medium with the constant pressure $p_0$.  The pressure within the nebula is determined by the energy injected by the equatorial striped wind, which transfers most of the energy of the flow. Since the residual magnetization of the striped wind zone is small, it is terminated at a strong shock. The equatorial radius of the termination shock, $a$, may be estimated from the pressure balance condition
\eqb
F_0=p_0a^2.
\label{equator_shock}\eqe
Inasmuch as the energy flux in the wind decreases with latitude, the shock is elongated. In all previous works, the termination shock was assumed to be strong everywhere; then the distance from the pulsar to the shock at high latitudes is estimated as $z=(1/2)\theta^2a$ \citep{Lyubarsky02}. Now I assume that
at $\theta<\theta_0$, the flow is highly magnetized and therefore it could be terminated only at a weak shock, which means that the postshock flow remains relativistic and radial. Then the shock should arise even closer to the pulsar in order to provide enough space for the flow to be adjusted to the external pressure.

The postshock flow could be matched to the external medium if it is causally connected. In fact the role of the termination shock is just to make the flow causally connected. It is quite possible that a single shock is unable to do the job; then a few shocks arise. For example, in weakly magnetized models, the high latitude flow passes two shocks: a (highly oblique) termination shock and then a rim shock \citep{Komissarov_Lyubarsky04,DelZanna04}. In any case, the flow eventually becomes causally connected. An important point is that for highly magnetized flow, it is not necessary to find the shape and number of the shocks because the postshock flow anyway remains radial so that one can look for the structure of the postshock flow just assuming that it is causally connected and at the inlet of the flow, the plasma moves radially.

Within the nebula, one can neglect the poloidal magnetic field as well as the azimuthal plasma velocity so the flow could be considered as composed from magnetic loops. Let the flow with the Lorentz factor $\gamma$ subtend the angle $\theta\gg 1/\gamma$. Then one can consider the flow as being composed from magnetic loops moving along the axis and expanding. In the frame moving together with the loop upwards with the velocity $c\cos\theta$, the loop expands relativistically with the Lorentz factor $\gamma_{\rm exp}=\theta\gamma$. The flow remains causally connected provided the expansion velocity exceeds the fast magnetosonic velocity. In high-$\sigma$ flows, the last is close to the speed of light, the corresponding Lorentz factor being $\gamma_{\rm fms}=\sqrt{\sigma}$. Then the causality condition, $\gamma_{\rm exp}<\gamma_{\rm fms}$, is written as
 \eqb
\theta\gamma<\sqrt{\sigma}.
 \label{causality}\eqe
Note that the flow could remain super-magnetospnic and even a highly super-magnetosonic if the flow opening angle is small. However, if the condition (\ref{causality}) is satisfied, the flow could be adjusted to the outer boundary conditions (e.g, \citet{bogovalov97}).
Below I assume that the injected radial flow already satisfies this condition. 

\section{Equations describing the high-latitude flow}
Let us for a while consider an axially symmetric flow.
One can find the flow equations from general asymptotic equations for axisymmetric flows \citep{lyubarsky09,Lyubarsky11}. However, it is not difficult to derive such equations straightforwardly taking into account from the beginning that the flow within the nebula is purely poloidal whereas the  magnetic field is purely toroidal so that one can write  
\eqb
\mathbf{v}=v\mathbf{l},\qquad \mathbf{B}=B\mathbf{e}_{\varphi}.
\label{kinematics}\eqe
Here $v$ is the flow velocity in units of $c$, $\mathbf{l}$ the unit vector along the flow
line, $\mathbf{e}_{\varphi}$ is the azimuthal unit vector. Now the condition of the flux freezing takes the form
 \eqb
\mathbf{E}=-\mathbf{v\times B}=-vB\mathbf{n},
 \label{flux_freez}\eqe
where $\mathbf{n=l\times e}_{\varphi}$ is the unit vector in the transverse direction.

In the steady state, one can introduce the electric potential
 \eqb
\mathbf{E}=-\nabla\Psi=-\vert\nabla\Psi\vert\mathbf{n}.
 \label{E}\eqe
The plasma flows along the equipotential surfaces therefore one can consider $\Psi$ as a specially normalized stream function; then the continuity equation is written as
 \eqb
2\pi r\rho\gamma v=\eta(\Psi)\vert\nabla\Psi\vert,
 \label{continuity}\eqe
where the function $\eta(\Psi)$ describes the distribution of the mass flux
at the inlet of the flow.

The  equation of motion of the hot postshock plasma is written in the steady state as
 \eqb
\rho\gamma(\mathbf{v\cdot\nabla})h\gamma\mathbf{v}=-\nabla p+
\frac 1{4\pi}\left[( \nabla\cdot\mathbf{E})\mathbf{E}+\mathbf{(\nabla\times B)\times B}\right];
\label{eq_motion} \eqe
where $\rho$ is the plasma proper density, $h$ specific enthalpy. One can conveniently make the projections of this equation on the direction of the flow, $\mathbf{l}$, and on the normal to the flow lines, $\mathbf{n}$.

The transfield force-balance equation may be obtained by taking the
dot product of Equation (\ref{eq_motion}) with $\mathbf{n}$. After some algebra (see, e.g., \citet{lyubarsky_eichler01}), we obtain in cylindrical coordinates
 \eqb
\frac 1{\cal
R}\left[h\rho\gamma^2v^2+\frac{E^2}{4\pi}\right]-\mathbf{n\cdot\nabla}p=
\frac1{8\pi r^2}\mathbf{\widehat n}\cdot\nabla
\left[r^2(B^2-E^2)\right];
 \label{transfield}\eqe
where $\cal R$ is the local curvature radius of the flow line (defined such that it is positive when the flow line is concave so that the flow is collimated towards the axis),
 \eqb
\frac 1{\cal R}= -\mathbf{n\cdot}(\mathbf{l\cdot\nabla})\mathbf{l}
=\mathbf{n\cdot}[\mathbf{l\times}(\mathbf{\nabla\times l})]=
-\mathbf{e}_{\varphi}\cdot(\mathbf{\nabla\times l}).
 \eqe

The flow velocity should be determined from the Bernoulli equation representing
the energy conservation along the flow line. This equation
may be obtained by taking the dot product of the momentum
equation (\ref{eq_motion}) with the longitudinal vector $\mathbf{l}$. However,
one can instead simply write down the conserved energy flux per unit mass flux
 \eqb
\frac{h\rho\gamma^2v+EB/4\pi}{\rho\gamma v}= \mu(\Psi);
 \eqe
which yields, with account of Eqs. (\ref{flux_freez}), (\ref{E}) and (\ref{continuity}),
 \eqb
h\gamma+\frac{r\vert\nabla\Psi\vert}{2\eta v}=\mu(\Psi).
 \label{Bernoulli}\eqe
 Note that the magnetization parameter
$\sigma$, defined as the ratio of the Poynting to the matter energy flux,
is presented  as
\eqb
\sigma=\frac{\mu-h\gamma}{h\gamma}.
\eqe

Now making use of Eqs. (\ref{flux_freez}), (\ref{E}) and (\ref{Bernoulli}) one reduces the transfield equation (\ref{transfield}) to the form
   \eqb
\frac{\eta\mu v\vert\nabla\Psi\vert}{2\pi r\cal
R}=\frac1{8\pi r^2}\mathbf{\widehat n}\cdot\nabla
\left(\frac{r^2\vert\nabla\Psi\vert^2}{\gamma^2-1}\right)+\mathbf{n\cdot\nabla}p;
 \label{transfield1}\eqe
In the relativistically hot medium, $p\propto\rho^{4/3}$, $h=4p/\rho$. With account of the   continuity and Bernoulli equations, one can write
 \eqb
\frac{p}{p_{\rm in}}=\left(\frac{r^2_{\rm in}\gamma_{\rm in}}{r^2\gamma}\frac{\mu-h\gamma}{\mu-h_{\rm in}\gamma_{\rm in}}\right)^{4/3}
 \label{pressure}\eqe
 \eqb
\frac{h}{h_{\rm in}}=\left(\frac{r^2_{\rm in}\gamma_{\rm in}}{r^2\gamma}\frac{\mu-h\gamma}{\mu-h_{\rm in}\gamma_{\rm in}}\right)^{1/3};
 \label{enthalpy}\eqe
where the index "in" is referred to the plasma parameters at the inlet of the flow.

Equations (\ref{Bernoulli}), (\ref{transfield1}), (\ref{pressure}) and (\ref{enthalpy}) form a complete set of equations. They should be complemented by appropriate boundary conditions. If the flow is confined by the pressure of the external medium, $p_{0}$, the pressure
balance condition should be satisfied at the boundary:
 \eqb
\frac 1{8\pi}(B^2-E^2)+p=p_{0}.
 \eqe
Making use of Eqs. (\ref{flux_freez}) and (\ref{E}), one writes the boundary condition
in the form
 \eqb
\left\{\frac{\vert\nabla\Psi\vert^2}{8\pi(\gamma^2-1)}+p\right\}_{\Psi=\Psi_0}=p_{0},
 \label{boundary} \eqe
where $\Psi_0$ is the potential of the last flow line. The boundary condition at the axis of the flow is just $\Psi(r=0)=0$.

The integrals of motion $\mu$ and $\eta$ are determined by the structure of the pulsar wind. The flow in the wind is radial so that $\Psi$, $\eta$ and $\mu$ depend only on the polar angle $\theta$.
Since the wind is highly relativistic, one can present the Poynting flux assuming $B=E=\vert\nabla\Psi\vert$, which yields
\eqb
F=\frac{EB}{4\pi}D^2=\frac 1{4\pi}\left(\frac{\partial\Psi}{\partial\theta}\right)^2.
\eqe
Here $D$ is the radial distance from the pulsar.
Comparing this expression with the angular distribution of the Poynting flux in the pulsar wind (\ref{energy_flux}), one finds the angular distribution of the electric potential in the wind:
\eqb
\Psi=\sqrt{4\pi F_0}(1-\cos\theta).
 \label{Psi_wind}\eqe
Within the wind, one can write  Equation (\ref{Bernoulli}) for the energy integral as (note that the wind is cold)
\eqb
\mu=\gamma_0+\frac{\sin\theta}{2\eta}\frac{\partial\Psi}{\partial\theta},
\eqe
where $\gamma_0$ is the Lorentz factor of the pulsar wind near the axis.
Eliminating $\theta$ with the aid of Equation (\ref{Psi_wind}), one can write $\mu(\Psi)$ close to the axis in the form
 \eqb
\mu(\Psi)=\gamma_0+\frac{\Psi}{\eta};
 \label{energy}\eqe
which represents in fact the first two terms in the universal expansion of $\mu(\Psi)$ in small $\Psi$ \citep{lyubarsky09}. The first term represents the plasma energy flux whereas the second the Poynting flux. Since the pulsar wind is Poynting dominated, the plasma term could play the role only extremely close to the axis. We are interested in the Poynting dominated domain of the wind, $\Psi\gg \eta\gamma_0$; therefore we will neglect the first term Equation (\ref{energy}).

In the Poynting dominated domain, $E^2/4\pi\gamma^2\gg h\rho=4p$, therefore one can also neglect the pressure term in the transfield equation (\ref{transfield1}) and in the boundary condition (\ref{boundary}). Since the flow is relativistic, one can also neglect unity as compared with $\gamma^2$ in the right-hand side and take $v=1$ in the left-hand side of the transfield equation. Now the transfield equation takes the form
   \eqb
\frac{\Psi\vert\nabla\Psi\vert}{\cal
R}=\frac1{4 r}\mathbf{\widehat n}\cdot\nabla
\left(\frac{r^2\vert\nabla\Psi\vert^2}{\gamma^2}\right).
 \label{transfield2}\eqe
 In the same approximation, Equations (\ref{pressure}) and (\ref{enthalpy}) are reduced to
 \eqb
h=h_{\rm in}\left(\frac p{p_{\rm in}}\right)^{1/4}= h_{\rm in}\left(\frac{r^2_{\rm in}\gamma_{\rm in}}{r^2\gamma}\right)^{1/3}.
 \label{state}\eqe

\section{The high-latitude flow; axisymmetric case}

In principle, the set of equations (\ref{Bernoulli}), (\ref{transfield2}) and (\ref{state}) could be reduced to a single equation for $\Psi$ just by expressing $\gamma$ via $\Psi$ from Equations (\ref{Bernoulli}) and (\ref{state}) and substituting the result into Equation (\ref{transfield2}). The problem is that in Poynting dominated flows, $\gamma$ is presented in the Bernoulli equation (\ref{Bernoulli}) as a small difference of two large terms, which makes the obtained equation for $\Psi$ inappropriate for approximate solution. One can circumvent this difficulty \citep{lyubarsky09} noticing that $\gamma$  could be easily found from the transfield equation (\ref{transfield2}) provided the shape of the magnetic surfaces, $\Psi(r,z)$, is known. An important point is that, in this case, an extra accuracy is generally not necessary because in the transfield equation, $\gamma$ is not presented as a difference of large terms. A special care should be taken only if the flow
becomes nearly radial because the curvature of the flux surfaces (the left-hand side of Equation (\ref{transfield2})) is determined in this case by small deviations of the flow lines from the straight lines.

Let us for a while neglect
corrections of the order $h\gamma/\mu$ (i.e. of the order of $1/\sigma$) to the shape of the flux line; validity of this approximation will be checked a posteriori. Then the Bernoulli Equation (\ref{Bernoulli}) is reduced, 
with account of Equation (\ref{energy}), to
 \eqb
r\vert\nabla\Psi\vert=2\Psi,
 \label{Bernoulli1}\eqe
which could be considered as an equation for $\Psi$.

One can find an analytical solution to the set of equations (\ref{transfield2}) and (\ref{Bernoulli1}) assuming that the flow subtends a small polar angle so that $z\gg r$. In this case one
can take $\mathbf{\widehat n}\cdot\nabla=\partial/\partial r$. When
looking for the shape of the magnetic surfaces, one can
conveniently use the unknown function $r(\Psi,z)$ instead of
$\Psi(r,z)$. Then, e.g.,
 \eqb
E=\vert\nabla\Psi\vert\approx\frac{\partial\Psi}{\partial r}=
\left(\frac{\partial r}{\partial\Psi}\right)^{-1}.
 \label{Bp_jet}\eqe
In the same approximation, the curvature radius may be presented as
(note that $\cal R$ is defined to be positive for concave surfaces)
 \eqb
\frac 1{\cal R}=-\frac{\partial^2r}{\partial z^2}.
 \label{curvature}\eqe
Now the transfield equation (\ref{transfield2}) and the Bernoulli equation (\ref{Bernoulli1}) are reduced to
  \eqb
-\Psi r\frac{\partial^2 r}{\partial z^2}=\frac{\partial}{\partial\Psi}
\left(\frac{\Psi^2}{\gamma^2}\right),
 \label{coll_transfield} \eqe
\eqb
2\Psi\frac{\partial r}{\partial\Psi}=r,
\label{Bernoulli0}\eqe
correspondingly.
With account of Equation (\ref{Bernoulli1}) the boundary condition (\ref{boundary}) takes the form
 \eqb
\frac{\Psi_0^2}{\left[r\gamma\right]^2_{\Psi=\Psi_0}}=2\pi p_{0},
 \label{boundary1} \eqe

The general solution to Equation (\ref{Bernoulli0}) may be presented as
 \eqb
r=R(z)\sqrt{\frac{\Psi}{\Psi_0}},
\label{r(psi)}\eqe
where $R(z)$ is an arbitrary function, which is in fact the cylindrical radius of the flow.
In order to find the function $R(z)$, let us substitute
Eq.(\ref{r(psi)}) into the left-hand side of Eq. (\ref{coll_transfield}) and integrate
the obtained equation from 0 to $\Psi_0$\footnote{Formally speaking, one could not integrate from 0 since the solution (\ref{r(psi)}) was obtained only in the Poynting dominated domain, where the first term in the full Bernoulli equation (\ref{Bernoulli}) is neglected. Therefore this solution becomes invalid close enough to the axis where the flow ceases to be Poynting dominated (the Poynting flux goes to zero at the axis, see Equation (\ref{energy})). However, the integrands grow with $\Psi$ so that the lower boundary condition could be continued to zero. For more details, see \citep{lyubarsky09} where a solution for the $\Psi\ll\Psi_0$ part of the flow, including the matter dominated core of the jet,  has been obtained; this solution is matched smoothly with the solution in the Poynting dominated domain.}:
 \eqb
-\frac{\Psi_0^2}3 R\frac{d^2R}{d z^2}=
\frac{\Psi_0^2}{\left(\gamma^2\right)_{\Psi=\Psi_0}}.
\eqe
Making use of the boundary condition (\ref{boundary1}) one gets a simple equation for $R$:
\eqb
\frac{d^2 R}{d z^2}+\beta R=0,
\label{govern}\eqe
where
\eqb
\beta=\frac{6\pi p_{0}}{\Psi_0^2}.
\eqe
Equation (\ref{govern}) is a partial case of the governing equation that
determines all the properties of highly magnetized flows \citep{lyubarsky09}.


Solution to Equation (\ref{govern}) describes the postshock flow at high latitudes where the wind remains highly magnetized.  Let the high magnetization flow subtend a small polar angle $\theta_0$. Then one can write, with account of Equations (\ref{equator_shock}) and (\ref{Psi_wind}),
\eqb
\beta=\frac{6}{\theta^4_0a^2}.
\eqe
Since the shock is weak, the flow just beyond the shock remains nearly radial so that if the flow line $\Psi=\Psi_0$ enters the shock at the distance $z_0$ from the pulsar, the solution to Equation (\ref{govern}) should satisfy the conditions: $R(z_0)=z_0\theta_0$; $R'(z_0)=\theta_0$. Such a solution is simply
 \eqb
R=\frac{\theta_0}{\sqrt{\beta}}\sin\sqrt{\beta}z=
\frac{\theta_0^3a}{\sqrt{6}}\sin\frac{\sqrt{6}\,z}{\theta_0^2a}.
 \label{flowshape}\eqe
Note that the solution is independent of the position of the shock, $z_0$. This is because the postshock flow remains anyway radial. At the end of this section, I present an estimate for $z_0$.

The Lorentz factor of the flow is obtained by substituting the solution (\ref{r(psi)}) and (\ref{flowshape}) into Equation (\ref{coll_transfield}) and integrating from zero to $\Psi$; this yields
 \eqb
\gamma=\sqrt{\frac{3\Psi_0}{\Psi}}\left(\theta_0\sin\frac{\sqrt{6}\,z}{\theta_0^2a}\right)^{-1}
 \label{gamma}\eqe
One sees that initially the flow expands and decelerates and then converges and accelerates. Substituting the obtained solutions into Equation (\ref{state}), one finds
 \eqb
h\propto \left(\sin\frac{\sqrt{6}\,z}{\theta_0^2a}\right)^{-1/3},
 \eqe
which means that the flow is cooled when it expands and heated again when it converges.

The flow is maximally expanded at the distance
\eqb
z_1=\frac{\pi}{2\sqrt{6}}\,\theta_0^2a.
\label{z1}\eqe
At this distance, the radius and the Lorentz factor of the flow are
\eqb
R_1=\frac{2}{\pi}\theta_0z_1=\frac{\theta^3_0}{\sqrt{6}}\,a;\qquad  \gamma_1=\frac{\sqrt{3}}{\theta_0}.
\eqe
Note that if the flow were weakly magnetized so that the termination shock were strong, the shock would arise at the distance $z=(1/2)\theta_0^2a\sim z_1$ \citep{Lyubarsky02}, roughly equal to $z_1$.
In  a highly magnetized flow, the shock is much closer to the pulsar.

The above solution was obtained by neglecting the $h\gamma$ term in the Bernoulli Equation (\ref{Bernoulli}). This means that the expression (\ref{r(psi)}) for the shape of the flow lines, where $R(z)$ satisfies the equation (\ref{govern}), is valid to within $h\gamma/\mu=1/\sigma\ll 1$. This approximation fails if the flow lines are nearly straight because in this case, the curvature of the flow lines (the left-hand side of the transfield Equation (\ref{coll_transfield})) is determined by small deviations from the straight line so that even small corrections to the flow line shape could not be neglected. In our solution, this happens near the origin of the flow, $z\approx 0$, and near the converging point, $z\approx 2z_1$.

In order to find limits of applicability of the solution, let us present the shape of the flow lines as (cf. Equation (\ref{r(psi)}))
\eqb
r=R(r)\sqrt{\frac{\Psi}{\Psi_0}}\,(1+\delta),
\eqe
where $\delta(\Psi,z)$ describes a small correction to the shape of the
flux surfaces due to a nonzero $h\gamma/\mu$. Substituting this expression into the full Bernoulli Equation (\ref{Bernoulli}) and expanding, with account of Equation (\ref{energy}) and (\ref{Bp_jet}), in small $\delta$, one gets
\eqb
h\gamma=\frac{2\Psi^2}{\eta}\frac{\partial\delta}{\partial\Psi},
\eqe
which implies $\delta\sim h\gamma/\mu=1/\sigma$.
One can neglect this correction provided the contribution to the curvature of the flow line due to this correction,
\eqb
\left\vert\frac{\partial^2R\delta}{\partial z^2}\right\vert\sim
\frac{R\delta}{z^2}\sim \frac R{\sigma z^2},
 \eqe
remains less than $d^2R/dz^2$. According to equation (\ref{govern}), the last is just equal to $-\beta R$ therefore the condition that one can neglect the corrections of the order of $1/\sigma$  to the shape of the flow line, which is in fact the validity condition for the solution described by Equations (\ref{r(psi)}) and (\ref{flowshape}), is written as
 \eqb
 z\gg \frac 1{\sqrt{\sigma\beta}}=\frac{\theta_0^2}{\sqrt{6\sigma}}a=\frac 2{\pi\sqrt{\sigma}}z_1.
 \label{causal_cond}\eqe
Taking into account that according to Equation (\ref{gamma}), the Lorentz factor of the flow at small $z$ is presented as $\gamma=\theta_0a/(\sqrt{2}\,z)$, one sees that this condition may be written as $\theta_0\gamma\ll\sqrt{\sigma}$, which is in fact the condition (\ref{causality}) of the causal connection of the flow. Therefore our solution is valid only in the domain of the causal connection of the flow.

The full structure of the flow could be formally obtained by matching the above solution with the free pulsar wind by inserting shocks and making use of the shock jump conditions.
I do not address here the entire  problem but just assume that shock(s) make the flow marginally causally connected, after which my solution is valid.
This conjecture could be justified by contradiction: if the flow remains causally disconnected (even having passed a shock), it could not be adjusted to the outer conditions and therefore more shocks must arise; if the flow is causally connected, it is smoothly adjusted to the outer boundary conditions according to the above solution. Therefore one could expect that the shock(s) just make the flow marginally connected, after which the above solution works. This means, according to the estimate (\ref{causal_cond}), that the shock(s) arise approximately at the distance
\eqb
z_0=\frac{\theta_0^2}{\sqrt{\sigma}}a\approx\frac{z_1}{\sqrt{\sigma}}.
\eqe

According to this solution, the flow is focused to a point $z=2z_1$ at the axis. Close enough to this point, just like close to the origin, the solution becomes invalid because the flow lines become nearly straight. But what seems to be more important is that such a converging flow ceases to be axisymmetric because of development of instabilities. This issue is discussed in the next section.

\section{Instabilities, turbulence and particle acceleration}

The axisymmetric flow considered in the previous section could not remain axisymmetric. The flow is composed from magnetic loops disconnected one from another therefore they could easily come apart. In the volume of the flow, the axisymmetry could be destroyed by the kink instability \citep{Begelman98,Mizuno11}. Near the boundary of the flow, the Kelvin-Helmholtz instability \citep{Begelman99} develops. In relativistic flows, these instabilities develop slowly because of relativistic time delay. However, if the flow converges, even small perturbations eventually destroy the regular structure. If two converging loops initially shifted one with respect to another by a displacement much less than their radius, the distortion becomes strong when the radius approaches the initial displacement. Therefore when the axixymmetric flow is focused into a point at the axis, the magnetic loops come apart close enough to the converging point so that a specific turbulence of shrinking magnetic loops emerges.

Taking into account that the hoop stress within the loop is not counterbalanced by either the poloidal magnetic field or the plasma pressure, the loops shrink with relativistic velocity, $\gamma\sim\sqrt{\sigma}$, until the plasma energy reaches the magnetic energy. Therefore one concludes that the energy of the high latitude, highly magnetized flow is released close to the converging point, $z=2z_1$. In this picture, the observed jet  \citep{Weisskopf00} begins in the vicinity of this point.  One can expect also an efficient particle acceleration  because in such a turbulence, the electric and magnetic fields are nearly equal.

The observed synchrotron spectrum from the Crab is extended up to $\sim 100$ MeV, which implies particle acceleration on a time scale of the Larmor gyration period so that the accelerating electric field should be as strong as the magnetic field \citep{Guilbert83,deJager96}. The recent discovery of daylong gamma ray flares in the band of a few hundred MeV \citep{Tavani11,Abdo11,Striani11,Vittorini11,Buehler12} poses even a larger challenge to the acceleration models \citep{Bednarek_Idec11,Komissarov_Lyutikov11,Uzdensky11,Bykov12,Cerutti12,Clausen_Lyutikov12,Sturrock_Aschwanden12}. The relativistic turbulence could in principle resolve the problem. On the one hand, at relativistic turbulent velocities or, which is the same, at $E\approx B$, even the second order Fermi process is efficient enough  to permanently accelerate electrons up to the energies sufficient to emit ~100 MeV synchrotron photons. On the other hand, random relativistic bulk motions could occasionally produce flares even in a harder energy band when a local radiating "knot" moves toward the observer \citep{Yuan11}.

The efficient particle acceleration could occur not only via Fermi mechanism but also via reconnection \citep{Uzdensky11,Cerutti12}. When different magnetic loops slide one over another, strong field gradients, and therefore strong currents, arise. The magnetic reconnection comes into play when the current velocity approaches the speed of light. The corresponding scale, $\delta\sim B/(8\pi en)$, could be expressed via the plasma multiplicity defined as the ratio of the plasma density to the Goldreich-Julian density in the pulsar magnetosphere, $\kappa= encP/B$, where $P$ is the pulsar period. Taking into account that within the magnetosphere, $n\propto B$, whereas beyond the light cylinder, $n\propto B/r$, one finds $\delta\sim r/\kappa$. This scale may be considered as the dissipation scale of the turbulence; at this scale the magnetic energy is converted into the plasma energy.

The detailed analysis of the particle acceleration is beyond the scope of this paper; the above consideration just shows that a plausible site for the synchrotron emission in the hundreds MeV band, and in particular for gamma-ray flares, is the region close to the point $z=2z_1$ toward which the high latitude flow is focused. According to Equation (\ref{z1}), this region is at the axis of the system at the distance a few times less than the equatorial radius of the termination shock. In this region, the energy transferred by the strongly magnetized high latitude flow is released. This region may be identified with the base of the observed jet \citep{Weisskopf00}.

\section{Conclusions}

In this paper, I considered the fate of the high latitude flow in the pulsar wind. This part of the wind remains strongly magnetized (see also \citet{Komissarov12}) therefore it is terminated at a weak shock (or at a system of weak shocks), beyond which the flow is still radial and relativistic. The shock arises very close to the pulsar, much closer than if it were strong; the higher the magnetization of the flow, the closer the shock to the pulsar. Beyond the shock, the flow still expands but decelerates and eventually becomes to converge because the magnetic hoop stress is not counterbalance either by the poloidal field or by the plasma pressure.

In the converging flow, magnetic energy is converted into the plasma energy therefore the plasma accelerates and heats. Even small perturbations due to, e.g., kink instability eventually destroy the converging flow so that magnetic loops come apart and then shrink independently of each other producing  relativistic turbulence. Hence one can expect that the whole energy of the highly magnetized part of the pulsar wind is released in a very small region close to the converging point, which occurs on the axis of the system at the distance from the pulsar $\sim\theta_0^2a$, where $\theta_0$ is the opening angle of the highly magnetized part of the wind, $a$ the equatorial radius of the termination shock. I identify this region with the base of the observed jet. Relativistic turbulent motions in highly magnetized plasma imply $E\approx B$ so that in the energy release region, particles could be efficiently accelerated either via the second order Fermi mechanism or via the magnetic reconnection. Therefore the synchrotron gamma-ray emission in the hundreds MeV band, both persistent and flaring, could come from a small region at the base of the jet.

\section*{Acknowledgments}
I am grateful to Serguei Komissarov and the anonymous referee for careful reading of the manuscript.
The work was supported by the Israeli Science Foundation under the grant 737/07.


 \end{document}